\documentstyle[prl,aps]{revtex}
\twocolumn
\begin{document}
\input{epsf}
\draft
\twocolumn[\hsize\textwidth\columnwidth\hsize\csname@twocolumnfalse\endcsname
\title{Normal scaling in globally conserved
interface-controlled coarsening of fractal clusters}
\author{Avner Peleg $^{1}$, Massimo Conti $^{2}$
and Baruch Meerson $^{1}$}
\address{$^{1}$ Racah Institute of Physics, Hebrew
University   of  Jerusalem, Jerusalem 91904, Israel}
\address{$^{2}$ Dipartimento di Matematica e Fisica,
Universit\'{a} di Camerino, and Istituto Nazionale di Fisica della
Materia, 62032, Camerino, Italy}
\maketitle
\begin{abstract}
Globally conserved
interface-controlled 
coarsening of fractal clusters exhibits dynamic
scale invariance and normal scaling. This is demonstrated 
by a numerical solution of
the
Ginzburg-Landau equation with a global conservation law. 
The sharp-interface limit of this equation is
volume preserving motion by mean curvature. The scaled form of the 
correlation 
function 
has a power-law tail accommodating the
fractal initial condition. The
coarsening length exhibits normal scaling with time. 
Finally, shrinking of
the fractal clusters with time is observed. The difference between global and 
local conservation is discussed.
\end{abstract}
\pacs{PACS numbers: 64.60.Ak, 61.43.Hv, 05.45.Df, 05.70.Fh}
\vskip1pc]
\narrowtext
Dynamics of growth of order from disorder in coarsening
systems with long-range
correlations
is an intriguing problem which appears in 
phase ordering \cite{Bray1} and in many other applications. 
Systems with long-range correlations are often characterizable by
fractal geometry \cite{Fractals,Vicsek}. Therefore, a lot of
attention has been devoted recently
to a variety of ``fractal coarsening" 
problems 
\cite{Toyoki,Jullien1,Irisawa1,Irisawa2,Jullien2,Meerson,Conti1,Conti2,Peleg,Crooks,Kalinin,Streitenberger}. 
A typical (though not the only)
setting for fractal coarsening
is the following. At an earlier stage of the dynamics
a fractal cluster (FC)
develops due to an
instability of growth of the ``minority 
phase" \cite{Vicsek,Meakin}. When the mass (or heat)
source is depleted, coarsening by
surface tension
becomes dominant. 
The main question concerns possible scaling relations and
universality classes in this type of
coarsening.

A major 
simplifying assumption 
in the analysis of a coarsening process is 
dynamic scale invariance (DSI) 
(see review \cite{Bray1}). 
The DSI hypothesis was first applied to fractal
coarsening by Toyoki and Honda \cite{Toyoki}
who considered systems with non-conserved order parameter. 
Implications  
of (mass) conservation in fractal coarsening
were considered more recently \cite{Jullien1,Meerson}. Most 
remarkable of 
them is predicted shrinking of the FCs in the process of 
coarsening. However, there has been no
convincing evidence (neither in experiment, nor in simulations)
in favor of DSI in conserved fractal coarsening. Moreover,
anomalous scaling and breakdown of DSI were observed
in recent simulations 
of
locally conserved edge-diffusion- \cite{Jullien2}
and bulk-diffusion-controlled \cite{Conti1,Conti2}
fractal coarsening.
This communication reports our finding 
that
DSI and normal scaling hold in the process of {\it interface-controlled}
fractal coarsening with a {\it globally} conserved 
order parameter. This system is apparently the first realistic
conserved fractal coarsening
system where this simplifying and beautiful concept is found to work.

Globally conserved interface-controlled coarsening
is accessible in experiment. Consider
the sublimation/deposition dynamics of a
solid and its vapor in a small closed vessel
kept at a (constant) low temperature \cite{airplane}. 
As the acoustic time in the gas phase is short compared
to the coarsening time, the gas pressure
(and, consequently, density) remains uniform in space, 
changing only in time. This character of mass 
transport
makes the coarsening dynamics conserved
globally rather than locally, which leads to a 
different kinetics.
Another example
appears in the context of attachment/detachment-controlled 
nanoscale fluctuations at solid surfaces 
\cite{Williams,Zinke-Allmang}. There is also a strong 
recent evidence in favor of interface-controlled transport
during the cluster coarsening in driven rapid granular 
flows \cite{Aranson1}. 

Here is an outline of the rest of the paper.
We shall work with
the Ginzburg-Landau equation with
a global conservation law. The corresponding
sharp interface
theory is reducible, at large times, to
volume preserving motion by mean curvature.
Assuming DSI, one can then
predict
scaling behavior
of the 
correlation function, 
shrinking of FCs with time and normal scaling
of the coarsening length $l(t)$. Our extensive numerical
simulations of the coarsening of two-dimensional (2d) 
diffusion-limited aggregates (DLAs)
support all these
predictions. We shall
conclude by pointing out the main difference between
global and local conservation.

We adopt a simple Landau
free energy functional:
\begin{equation}
F[u]=\int\left[(1/2) ({\bf{\nabla}}u)^2+V(u)+Hu\right]\,d^{d}{\bf r}
\,,
\label{1}
\end{equation}
where $V(u)=(1/4) (1-u^{2})^{2}$ is a double-well potential, 
$u({\bf r},t)$ is the order parameter 
and fluctuations are neglected. 
The effective ``magnetic field" $H=H(t)$ 
changes in time
so as to impose a global conservation law: $<u>=\mbox{const}$,
where $<...>$ denotes a spatial average:
\begin{equation}
<...>= L^{-d}\,\int(...)\, d^{d}{\bf r}
\,,
\label{2}
\end{equation}
$L$ is the system size and the integration is
over the whole system. The dynamics is described by a simple
gradient descent:
\begin{equation}
\frac{\partial u}{\partial t}=-\frac{\delta F}{\delta u}= \nabla
^{2}u+u-u^{3}-H(t)
\,.
\label{3}
\end{equation}
It follows from Eq. (\ref{3}) and the conservation law that
$H(t)=<u-u^{3}>$ (either no-flux, or periodic boundary conditions
are assumed), so Eq. (\ref{3}) is a nonlocal
reaction-diffusion equation \cite{Schimansky,Rubinstein,Mikhailov,MS}.
In the context of phase ordering it can be called
the Ginzburg-Landau equation with a global conservation law.
To make a theoretical progress, one should
work in the sharp-interface limit \cite{MS}
valid at late times, when the system already consists of large domains 
of ``phase 1"
and ``phase 2" divided by a sharp interface.
At this stage $H(t)$ is both 
small, $H(t) \ll 1$, and slowly varying in time. 
The phase 
field in the phases $1$ and $2$ is uniform and rapidly adjusts to the 
current value of $H(t)$, so $u=-1-H(t)/2$
and $1-H(t)/2$, respectively. For brevity, we will consider
2d-case. The normal velocity of the
interface is
\begin{equation}
v_{n}(s,t)=\frac{3}{\sqrt{2}}H(t)-\kappa(s,t)
\,,
\label{8}
\end{equation}
where $s$ is the coordinate along the interface and $\kappa$ is the
local curvature of the interface. The positive sign of $v_n$
corresponds to the interface moving toward phase $2$, while
$\kappa$ is positive when the interface is
convex towards phase $2$.

An equation for $H(t)$ follows from the global
conservation law which takes the form
\begin{equation}
\frac{4 A(t)}{L^{2}}-H(t)=\mbox{const}
\,,
\label{11}
\end{equation}
where 
$A(t)=\int_{u({\bf r},t)>0}d^{2}{\bf r}$
is the cluster area.

Eqs. (\ref{8}) and (\ref{11}) make a closed set and provide
a general sharp-interface formulation to the Ginzburg-Landau
equation with a global conservation law. Often
they
can be 
simplified further. Compute
the area loss rate of the cluster:
\begin{equation}
\dot{A}(t) = \oint v_{n}(s,t)ds = \Lambda(t) \left[
\frac{3}{\sqrt{2}}H(t)-\overline{\kappa(s,t)}\right]
\,, \label{12}
\end{equation}
where $\overline{\kappa(s,t)}$ is the curvature averaged over the whole
interface:
\begin{equation}
\overline{\kappa(s,t)}=\frac{1}{\Lambda(t)}\oint \kappa(s,t)ds
 \,,
\label{13}
\end{equation}
and $\Lambda(t)$ is the cluster perimeter. If the cluster area is conserved, 
then $H(t) = (\sqrt{2}/3)\overline{\kappa(s,t)}$ which yields
\begin{equation}
v_{n}(s,t)=\overline{\kappa(s,t)}-\kappa(s,t)
 \,.
\label{14}
\end{equation}
This is area-preserving motion by curvature (in 2d), or
volume-preserving motion by mean curvature (in 3d)
\cite{Rubinstein,MS,Gage}. 
Dynamics (\ref{14}) shortens the interface
length (in 2d), or area (in 3d) 
\cite{Rubinstein,Gage}. This nonlocal
coarsening model
is simpler than the better known
``Laplacian coarsening model" (derivable 
from
the Cahn-Hilliard equation \cite{Bray1,Pego}) which describes
the late-time asymptotics of the
locally-conserved bulk-diffusion-controlled coarsening.

Assuming DSI and using Eq.
(\ref{14}), one obtains 
normal scaling for 
the characteristic coarsening length:
$l(t)\sim t^{1/2}$. Therefore,
global conservation does not change the scaling law.
The same result (again, when assuming DSI)
follows from dynamic renormalization group 
arguments applied to Eq. (\ref{3}) (with a
Gaussian white noise term) \cite{Bray2}. 
For short-range correlations 
this result was supported by
particle simulations of critical \cite{simulations} and off-critical
\cite{Majumdar} quench, and by
a numerical solution of Eq. (\ref{3}) for both critical, and off-critical
quench \cite{CMPS}.
 
Let us return to fractal coarsening. The initial conditions 
represent FCs that are
characterizable by their fractal dimension $D$ on an
interval of scales between the lower cutoff $\tilde{l}_{0}$ and
the upper cutoff $\tilde{L}_{0}$. The DSI-based coarsening
scenario \cite{Toyoki,Jullien1,Meerson} assumes that the fractal
dimension of the cluster remains constant on a
shrinking interval of distances between the lower cutoff
$\tilde{l}(t)$ (which has the same dynamic scaling as the 
coarsening length), and the upper cutoff $\tilde{L}(t)$. Now, the
perimeter $\Lambda$ and area $A$ of the FC can be estimated
as \cite {Fractals}
\begin{equation}
\Lambda \sim l(\tilde{L}/l)^{D} \;\;\; \mbox{and} \;\;\; A \sim
l^{2}(\tilde{L}/l)^{D}
 \,,
\label{15}
\end{equation}
respectively. Then area conservation yields $L\sim
l^{(D-2)/D} \sim t^{-(2-D)/2D}$ which implies that the FC
shrinks with time \cite{Jullien1,Meerson}. This follows
$\Lambda(t)
\sim l^{-1}(t) \sim t^{-1/2}$. One can also predict the
asymptotic shape of the equal-time pair 
correlation function at large times: $C(r,t) \rightarrow g[r/l(t)]$.
At
distances $r\ll l(t)$ from a typical reference point inside the
cluster the correlation function should obey the Porod law:
$g(\xi)= 1-k \xi$
with a constant $k$ of order unity. At 
$l(t) \ll r \ll \tilde{L}(t)$
$g (\xi) \sim \xi^{D-2}$ (see Ref. \onlinecite{Meerson}), a
power-law tail with the same exponent as in $C(r,t=0)$.

In order to check these predictions, we solved
Eq. (\ref{3}) numerically on a domain
$2048 \times 2048$ with no-flux boundary conditions. 
The accuracy of the numerical scheme was monitored by
checking the (approximate) conservation law (\ref{11}) which 
was found to hold
with an accuracy better than $0.2${\%}
for $t>3$.

We used $10$ different
DLA clusters \cite {Witten} as the initial conditions.
These clusters (like the one shown
in Fig. 1, upper left) had a radius of order $10^3$.
To prevent fragmentation at an early stage of coarsening, 
the clusters were reinforced by an addition of peripheral
sites, similar to Ref. \onlinecite{Irisawa1}. The 
average fractal dimension of the
initial clusters, determined from the averaged pair correlation
function, was $1.75$.

Introducing the density $\rho({\bf
r},t)=(1/2) [u({\bf r},t)+1]$, we identified 
the cluster as the locus where 
$\rho({\bf r},t)\geq 1/2$. Typical snapshots of the
coarsening process are shown in Fig. 1. One can see that
larger features of the FC grow at the expense of smaller ones. 
At late times the cluster shrinks, in
agreement with the prediction of DSI \cite{shrinking}. A similar
shrinking is evident in the pictures obtained in Monte Carlo simulations
of area-preserving
interface-controlled coarsening \cite{Irisawa2}, although the 
authors of Ref. \onlinecite{Irisawa2} did not comment on it.

To characterize the coarsening process, several quantities were
sampled and averaged over the $10$ initial conditions:
\begin{enumerate}
\item The cluster area.
\item The (circularly averaged) correlation function,
normalized at r=0:
\begin{equation}
C(r,t)=\frac{<\rho({\bf r'}+{\bf r},t)\rho({\bf
r'},t)>}{<\rho^{2}({\bf r'},t)>} \,. \label{31}
\end{equation}
\item The coarsening length scale $l(t)$, computed
from the equation $C(l,t)=1/2$. 
$l(t)$ can be
interpreted as the typical width of the cluster branches.
\item The cluster perimeter $\Lambda (t)$ computed
by a standard 
algorithm \cite {Parker}.
\end{enumerate}
\begin{figure}[h]
\hspace{-1.0cm}
\rightline{ \epsfxsize = 6.5cm \epsffile{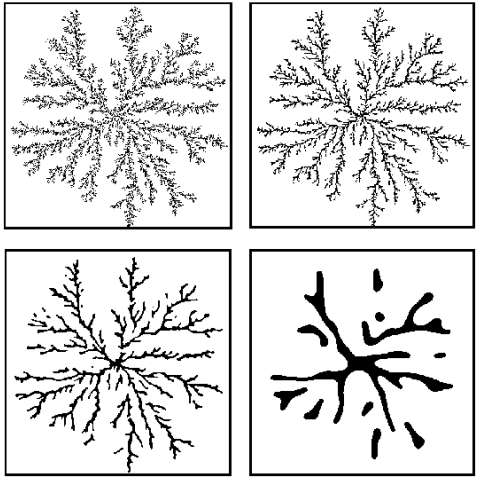}}
\vspace{0.1in}
\caption{
Evolution of a DLA cluster undergoing an
interface-controlled coarsening in a
globally conserved system. The upper row
corresponds to $t=0$ (left) and $12.6$ (right), the lower row to
$t=126.4$ (left) and $1856.6$ (right).
\label{fig. 1}}
\end{figure}

The cluster area was found to be constant with an accuracy
better than $0.5{\%}$ for $t>10$, and better than $0.15{\%}$ for
$t>100$. Hence, area preserving motion by curvature, Eq. (\ref{14}),
provides an accurate description to this regime. Figure 2 shows that,
at late times ($t>100$),
$C(r,t)$ approaches a scaled form. The scaled function
has a long-range power-law tail with an exponent $D-2$ 
(the same as in the initial condition), see the inset
of Fig. 2. Noticeable is the absence of any
additional dynamic length scales, in a striking 
contrast to the locally conserved
fractal coarsening \cite{Conti2}. The dynamics of
$l(t)$ is shown in Fig. 3.
The same figure shows a pure $t^{1/2}$ power-law line
(serving as a
reference for the expected late-time
behavior) and a corrected power-law fit
$l(t)=l_{0}+bt^{\alpha}$ with $\alpha =0.49$, $b=1.2$
and $l_{0}=5.0$. 

\begin{figure}[h]
\hspace{-2.6cm}
\rightline{ \epsfxsize = 6.0cm \epsffile{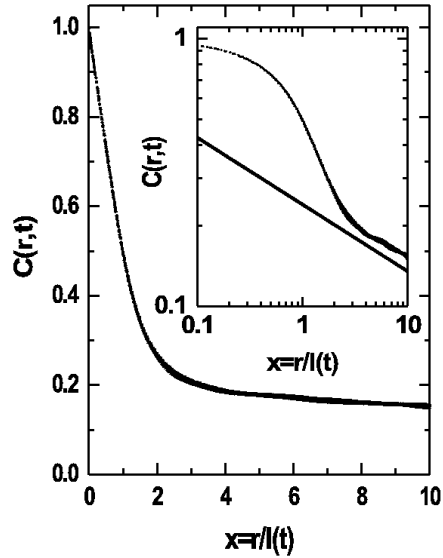}}
\vspace{0.1in} 
\caption{Scaling form of the
correlation function $C(r,t)$ for time moments
$t=400.0$, 587.0, 1264.8 and
1856.6. The inset shows the same data on a log-log
plot. The solid line, serving as a reference, represents a
power-law with an exponent $D - 2 = - 0.25$.
\label{fig. 2}}
\end{figure}

\begin{figure}[h]
\hspace{-2.6cm}
\rightline{ \epsfxsize = 6.0cm \epsffile{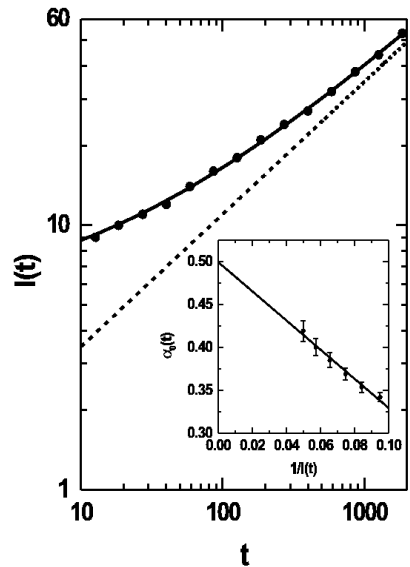}}
\vspace{0.1in} 
\caption{The coarsening
length $l(t)$ versus time (circles). The solid line is a
corrected power-law fit: $l(t)=l_{0}+bt^{\alpha}$ with
$\alpha=0.49$, $l_{0}=5.0$ and $b=1.2$. The dotted line represents a pure
$t^{1/2}$ power law. The inset shows the time-dependent
effective dynamic exponent $\alpha_{0}(t)$ versus
$1/l(t)$. The solid line is a linear fit. 
\label{fig. 3}}
\end{figure}

The inset of Fig. 3 illustrates a different method \cite{Huse} 
of determining the dynamic exponent. One defines
a (time-dependent) effective exponent:
$\alpha_{0}(t)=d \ln l(t)/d \ln t$. Under the normal scaling 
assumption, one can determine the
``true" dynamic exponent by plotting $\alpha_{0}(t)$ versus 
$l^{-1}(t)$ and extrapolating it to $t \to \infty$, that is 
to $l^{-1}(t) \to 0$. The numerical value of $\alpha_{0}(t)$ is computed
from
$\alpha_{0}(t)=\log_{10}\left[l(10t)/l(t)\right]$.
This procedure yields $\alpha = 0.50$. Therefore, $l(t)$ exhibits normal
scaling.

The same procedures were used for an analysis of the dynamic 
behavior of the cluster perimeter
$\Lambda (t)$. We found that $\Lambda^{-1}(t)$ exhibits the same normal
scaling: $\Lambda^{-1}(t)\sim t^{1/2}$. Irisawa {\it et al.}
\cite{Irisawa2}
reported an exponent $0.38$ for the inverse perimeter.
Their graph shows, however, that the 
effective
exponent increases at late times. We believe that a 
careful analysis of their data
would also lead to an exponent of $1/2$.

Thus, all predictions following from the DSI hypothesis: the normal
scaling of $l(t)$, shrinking of the FC and scaling behavior
of correlation function (including its
power-law tail), are confirmed by 
numerical simulations. We therefore
conclude that globally-conserved interface-controlled fractal
coarsening
exhibits DSI and normal scaling. This
behavior stands in contrast to the breakdown of
scale invariance observed in diffusion-controlled coarsening of
FCs \cite {Conti1,Conti2}, where the order
parameter is conserved {\it locally}. The mechanism of scaling violations
in locally-conserved systems is not entirely
clear at present, therefore
a comparison between the two types of systems can be instructive.
We relate the difference in scaling behavior 
to an important (and simple) difference in the character of transport. 
Global
transport, characteristic for interface-controlled
systems, is uninhibited by Laplacian screening effects typical for 
locally conserved systems.
Therefore, large-scale dynamics is always present in globally-conserved
systems, but is suppressed in locally-conserved ones.
This difference is observed already in a very
simpler setting of an area-preserving coarsening (shrinking) 
of long slender
bars. In the locally-conserved case the
bar acquires a dumbbell shape, while its initial width remains
(almost) constant and
represents a relevant length scale until late times
\cite{Conti2}. On the contrary, in the globally
conserved case the shrinking bar 
has a finger-like shape, and its dimensions are changing on
the same time scale \cite {Meerson2}.

We are grateful to Azi Lipshtat for
help. This work was supported in part by a grant from the Israel
Science Foundation, administered by the Israel Academy of Sciences and
Humanities.


\begin{references}
\bibitem{Bray1} A.J. Bray, Adv. Phys. {\bf 43}, 357 (1994).
\bibitem{Fractals} B.B. Mandelbrot, {\it The Fractal Geometry of
Nature} (Freeman, San Francisco, 1982); J. Feder, {\it Fractals}
(Plenum, New York, 1988).
\bibitem{Vicsek} T. Vicsek, {\it Fractal Growth Phenomena}
(World Scientific, Singapore, 1992).
\bibitem{Toyoki}H. Toyoki and K. Honda, Phys. Lett. {\bf 111A},
367 (1985).
\bibitem{Jullien1} R. Semp\'{e}r\'{e}, D. Bourret, T. Woignier, J.
Phalippou, and R. Jullien, Phys. Rev. Lett. {\bf 71}, 3307 (1993).
\bibitem{Irisawa1} T. Irisawa, M. Uwaha, and Y. Saito, Europhys.
Lett. {\bf 30}, 139 (1995).
\bibitem{Irisawa2} T. Irisawa, M. Uwaha, and Y. Saito, Fractals {\bf 4}
251 (1996).
\bibitem{Jullien2} N. Olivi-Tran, R. Thouy and R. Jullien, J.
Phys. I {\bf 6}, 557 (1996).
\bibitem{Meerson}B. Meerson and P.V. Sasorov,
cond-mat/9708036.
\bibitem{Conti1} M. Conti, B. Meerson, and P.V. Sasorov, Phys. Rev.
Lett. {\bf 80} 4693 (1998).
\bibitem{Conti2} M. Conti, B. Meerson, and P.V. Sasorov,
cond-mat/9912426.
\bibitem{Peleg} A. Peleg and B. Meerson, Phys. Rev. E {\bf 59}, 1238 (1999);
{\bf 62} (2000).
\bibitem{Crooks} G.E. Crooks, B. Ostrovsky, and Y. Bar-Yam, Phys.
Rev. E {\bf 60}, 4559 (1999).
\bibitem{Kalinin} S.V. Kalinin {\it et al.}, Phys. Rev. E {\bf 61}, 
1189 (2000).
\bibitem{Streitenberger} P. Streitenberger, in {\it Paradigms of Complexity.
Fractals and Structures in Sciences}, edited by M.N. Novak (World
Scientific, Singapore, 2000), p. 135.
\bibitem{Meakin} P. Meakin, {\it Fractals, Scaling and Growth Far
from Equilibrium} (Cambridge University Press, Cambridge, 1997).
\bibitem{airplane} One can observe an uncontrolled version of
this phenomenon while 
looking at slowly changing ice patterns on
the double window of an airplane flying at a high altitude. 
\bibitem{Williams} Z. Toroczkai and E. Williams,
Phys. Today {\bf 52} (12), 24 (1999).
\bibitem{Zinke-Allmang} M. Zinke-Allmang, L.C. Feldman, and
M.H. Grabow, Surface Sci. Reports {\bf 16}, 277 (1992).
\bibitem{Aranson1} I. S. Aranson {\it et al.}, Phys. Rev. Lett. {\bf 84}, 
3306 (2000). 
\bibitem{Schimansky} L. Schimansky-Geier, Ch. Z\"{u}licke,
and E. Sch\"{o}ll, Z. Phys. B {\bf 84}, 433 (1991).
\bibitem{Rubinstein} J. Rubinstein and P. Sternberg, IMA J. Appl.
Math. {\bf 48}, 249 (1992).
\bibitem{Mikhailov}A.S. Mikhailov, {\it Foundations of Synergetics
I. Distributed Active Systems} (Springer-Verlag, Berlin, 1993).
\bibitem{MS} B. Meerson and P.V. Sasorov, Phys. Rev. E {\bf 53}, 3491
(1996).
\bibitem{Gage} M. Gage, Contemporary Math. {\bf 51}, 51 (1986).
\bibitem{Pego} R.L. Pego, Proc. R. Soc. Lond. A {\bf 422}, 261 (1989).
\bibitem{Bray2} A.J. Bray, Phys. Rev. Lett. {\bf 66}, 2048 (1991),
and references therein.
\bibitem{simulations} J.F. Annett and J.R. Banavar, Phys. Rev. Lett. 
{\bf 68}, 2941 (1992); L.L. Moseley, P.W. Gibbs, and N. Jan, J. 
Stat. Phys. {\bf 67}, 813 ( 1992); A.D. Rutenberg, Phys. Rev. 
E {\bf 54}, 972 (1996).
\bibitem{Majumdar} C. Sire and S.N. Majumdar, Phys. Rev. E {\bf 52}, 244
(1995).
\bibitem{CMPS} M. Conti, B. Meerson, A. Peleg, and P.V. Sasorov,
in preparation.
\bibitem{Witten} T.A. Witten, Jr. and L.M. Sander, Phys. Rev.
Lett. {\bf 47}, 1400 (1981).
\bibitem{shrinking} The predicted ``shrinking exponent" 
$(D-2)/(2 D) \simeq - 0.07 $ is too small to be measured accurately. 
\bibitem{Parker} J.R. Parker, {\it Practical Computer Vision Using C}
(Wiley, New York, 1993), p. 51.
\bibitem{Huse} D.A. Huse, Phys. Rev. B, {\bf 34}, 7845 (1986).
\bibitem{Meerson2} A. Peleg, M. Conti,
B. Meerson, and A.J. Vilenkin, in preparation.
\end{references}
\end{document}